# Pro-arrhythmogenic effects of heterogeneous tissue curvature: A suggestion for role of left atrial appendage in atrial fibrillation

Jun-Seop Song[1]; Jaehyeok Kim[1]; Byounghyun Lim[1]; Young-Seon Lee[1]; Minki Hwang[1]; Boyoung Joung[1]; Eun Bo Shim[2]; Hui-Nam Pak[1,*]

[1]*Yonsei University Health System, Seoul, Republic of Korea*

[2]*Department of Mechanical and Biomedical Engineering, Kangwon National University, Chuncheon, Republic of Korea*

Short title: Arrhythmogenic role of tissue curvature

Address for correspondence:
    Hui-Nam Pak, MD, PhD
    50 Yonseiro, Seodaemun-gu,
    Seoul, Republic of Korea 120-752
    Phone: +82-2-2228-8460
    Fax: +82-2-393-2041
    E-mail: hnpak@yuhs.ac

# Abstract

**Background** The arrhythmogenic role of atrial complex morphology has not yet been clearly elucidated. We hypothesized that bumpy tissue geometry can induce action potential duration (APD) dispersion and wavebreak in atrial fibrillation (AF).

**Methods and Results** We simulated 2D-bumpy atrial model by varying the degree of bumpiness, and 3D-left atrial (LA) models integrated by LA computed tomographic (CT) images taken from 14 patients with persistent AF. We also analyzed wave-dynamic parameters with bipolar electrograms during AF and compared them with LA-CT geometry in 30 patients with persistent AF. In 2D-bumpy model, APD dispersion increased (p<0.001) and wavebreak occurred spontaneously when the surface bumpiness was higher, showing phase transition-like behavior (p<0.001). Bumpiness gradient 2D-model showed that spiral wave drifted in the direction of higher bumpiness, and phase singularity (PS) points were mostly located in areas with higher bumpiness. In 3D-LA model, PS density was higher in LA appendage (LAA) compared to other LA parts (p<0.05). In 30 persistent AF patients, the surface bumpiness of LAA was 5.8-times that of other LA parts (p<0.001), and exceeded critical bumpiness to induce wavebreak. Wave dynamics complexity parameters were consistently dominant in LAA (p<0.001).

**Conclusion** The bumpy tissue geometry promotes APD dispersion, wavebreak, and spiral wave drift in *in silico* human atrial tissue, and corresponds to clinical electro-anatomical maps.

# Introduction

Atrial fibrillation (AF) is one of the most common cardiac arrhythmias that increases the risk of cardioembolic stroke and mortality. Although pulmonary vein isolation (PVI) is an effective treatment for AF, PVI-only ablation is not sufficient particularly in patients with persistent AF.[1] In 2010, Di Biase et al. reported that the left atrial appendage (LAA) was found to be responsible for atrial arrhythmias in 27% of patients undergoing redo ablation.[2] Several other clinical studies support that the LAA is a potential extrapulmonary source of atrial arrhythmias and LAA isolation is effective for the treatment of persistent AF.[3-6] However, how the LAA play a pro-arrhythmic role in the maintenance of AF has not been clearly revealed.

One of the significant features of the LAA is its complex bumpy morphology.[7, 8] Several computational and experimental studies have shown that the uneven tissue geometry induce dispersion of action potential duration (APD)[9, 10] and change spiral wave dynamics.[8, 11, 12] Rogers[13] reported that nonzero Gaussian curvature (briefly, 'curvature') of tissue geometry altered wave propagation speed and APD, and abrupt change of curvature could promote wavebreak in ventricular geometry. Despite being studied previously, the mechanism and arrhythmogenic role of the bumpy atrial tissue have not been clearly established in human AF.

We hypothesized that bumpy tissue geometry could play a pro-arrhythmic role in AF. We studied the effects of heterogeneous curvature on spatial dispersion of APD and wavebreak in *in silico* human atrial tissue by varying the degree of bumpiness. To determine the pro-arrhythmic effects of bumpy feature in real LA geometry, we simulated 3D-left atrial (LA) model reflecting the LA geometry of AF patients. Additionally, we measured surface bumpiness of the LAA, and estimated wave dynamics parameters from clinically obtained bipolar electrogram, such as complex fractionated atrial electrogram (CFAE) and entropy parameters (Shannon and approximate entropy), which are known to be related to wavebreak[14] and rotational activation[15, 16].

# Methods

*Computational model of bumpy surface*

We simulated homogeneous, isotropic bumpy tissue consisting of 512×512 human atrial cells (Δx=0.025 cm). We designed periodically bumpy patterns by varying the degree of bumpiness as the following equation:

$$z = A\sin\left(\frac{2\pi kx}{L}\right)\cos\left(\frac{2\pi ky}{L}\right)$$

where L=12.8 cm is the size of the tissue, x, y are coordinates of the cells (0≤x, y≤L), A (mm) is the amplitude of curves, and k (no unit) is the number of periodic curves. We constructed 24 surfaces by varying the amplitude of curves A=1~6 mm and the number of periodic curves k=1~4 (Fig. S1-A). We considered the flat tissue as the control case.

A biophysical model of the human atrial myocyte[17] was implemented. No electrical remodeling was assumed to determine the pure effect of bumpy geometry on wavebreak. We calculated electrical conduction by the following reaction-diffusion equation:

$$\frac{\partial V}{\partial t} = -\frac{I_{ion} + I_{stim}}{C_m} + D\nabla^2 V$$

where V (mV) is the transmembrane potential, $C_m$=100 pF is the capacitance of myocyte, $I_{ion}$ and $I_{stim}$ (pA) are the total transmembrane currents and stimulus current, respectively, D=0.001 cm$^2$/ms is the diffusion coefficient. The diffusion term $\nabla^2 V$ was treated as the Laplace-Beltrami operator $\nabla^2 V = \frac{1}{\sqrt{g}}\sum_{i,j}\partial_i(\sqrt{g}g^{ij}\partial_j V)$ where $g^{ij}$ was inverse of the metric $g_{ij}$ and g was the determinant of the metric $g_{ij}$ (see Supplementary information).

To study spatial dispersion of APD in bumpy geometry, we measured APD map at steady state pacing. A vertical line located at the left side of the tissue was stimulated at 2 Hz, and APD was measured across the grid at the tenth beat. APD was calculated as the time interval between the time of maximum dV/dt and repolarization to -60mV, which is related to effective refractory period.[18] The degree of APD dispersion was estimated by standard deviation (SD) and range (max-min) of the APD. We excluded the boundary region defined by 5% of the tissue size in the analysis to eliminate the effects of tissue boundary.

We initiated spiral wave using a S1S2 cross-field protocol. A vertical line was stimulated (S1), and further, horizontal half plane was stimulated (S2) when recovery front of S1 stimulus reached half of the tissue. After initiation of the spiral wave, we examined whether wavebreak spontaneously occurs for 5 seconds. Additionally, we re-performed all the simulations after standardizing the area of the surfaces.

In the above part, we designed the surfaces with uniformly bumpy patterns, however, real LA geometry containing both flat and complex bumpy structure has non-uniformly surface bumpiness. To examine how heterogeneous bumpiness of atrial tissue affects spiral wave dynamics, we simulated 200 random Gaussian bumpy surfaces[19] and estimated wavebreak generation and AF sustainability defined as maintenance duration > 5 s. Furthermore, we also applied the bumpiness gradient and/or the ionic current gradient to determine the effects of heterogeneous bumpiness on spiral wave drift. A detailed protocol was described in Supplementary information.

### *Simulation of AF on 3D human LA model*

We simulated an *in silico* human LA model reflecting the LA anatomy of 14 AF patients (Table S1) who underwent radiofrequency catheter ablation (RFCA), as previously described by Hwang et al[20]. The study protocol adhered to the Declaration of Helsinki and was approved by the Institutional Review Board of Yonsei University Health System. The written informed consent was obtained for the use of cardiac CT and RFCA data (clinicaltrials.gov; NCT02171364). Ionic current was calculated using human atrial myocyte model[17] and electrical wave conduction was modeled by the mono-domain reaction-diffusion equation on the LA geometry (CUVIA, Model: SH01, ver. 1.0, Laonmed Inc.). We incorporated atlas-based fiber orientation to represent anisotropic electrical conduction. After the induction of AF, we calculated the density of phase singular (PS) points [21] by dividing the number of PS points by the total number of nodes during 6 seconds of AF. A detailed protocol was described in Supplementary information.

*Clinical electrophysiological mapping of wave dynamics parameters*

We analyzed clinically obtained bipolar electrogram signals to evaluate electrical wave dynamics of AF. Electrophysiological mapping was performed during sustained AF in 30 patients with persistent AF (22 men, 62.2±11.7 years old, Table S1) who underwent RFCA (see Supplementary information). The CFAE-cycle length (CL) was calculated by taking the average of time intervals between consecutive deflections. The deflections were identified as downstroke morphology between the local-maximum and the local-minimum amplitudes within a time duration of 15 ms which was set to avoid detection of far-field events. Additionally, we set a refractory period of 40 ms to avoid multiple detections of a single deflection event. CFAE was defined as CFAE-CL ≤ 120 ms.

We constructed Shannon entropy (ShEn) map as described by Ganesan et al[15]. Amplitude of bipolar signal was binned into voltage histogram (0.01 mV fixed bin size). The relative probability density $P_i$ was calculated by dividing the number of counts in *i*th bin by the sum of counts in all bins. The ShEn (no unit) was calculated as:

$$\text{ShEn} = -\sum_{i=0}^{N-1} P_i \log_2 P_i$$

where N is the total number of bins.

Additionally, we calculated approximate entropy (ApEn) map to quantify regularity of the bipolar signals[22]. We used the standard parameters with embedding dimension $m = 2$ and threshold $r = 0.1$ as described by Orozco-Duque et al[16].

*Calculation of curvature and surface bumpiness*

We calculated the curvature of tissue geometry at each node of the surface mesh[23]. To represent the degree of bumpiness of the surface, we defined "surface bumpiness" by SD of the curvature as follows:

$$\text{Surface bumpiness} = \sqrt{\frac{1}{S}\iint (K - \bar{K})^2 \, dA}$$

where S (cm$^2$) is the surface area, K (cm$^{-2}$) is the curvature, and $\bar{K}$ (cm$^{-2}$) is the average of the curvature on the surface. The surface bumpiness (cm$^{-2}$) represents the heterogeneity of the curvature on the surface. The pulmonary veins and the mitral valve were excluded from the analysis.

*Statistical analysis*

All data were presented as mean±SD. Surface bumpiness and wave dynamics parameters were compared between the LAA and the other LA parts using a paired t-test. Linear correlation analysis was performed to determine the relationship between two continuous variables. All statistical analysis of small samples (n<30) were performed with nonparametric tests. The relationship between degree of APD dispersion and surface bumpiness was tested with Spearman correlation analysis. A p-value<0.05 was considered to be statistically significant.

# Results

**Spiral wave dynamics in 2D-bumpy surface**

*Dispersion of APD depending on curvature*

In the 2D bumpy atrial model, we measured APD map in the steady state pacing. There was heterogeneity of APD induced by bumpy geometry in all the 24 bumpy surfaces. As shown in the flat control case (Fig. 1A), APD was significantly changed in stimulation site and exit site despite the zero-curvature of the tissue. This phenomenon is known as 'edge effect'.[9] To estimate the pure effect of curvature on APD dispersion, we excluded the boundary region in the analysis. In all the 24 bumpy surfaces, there was a linear correlation between APD and curvature (the mean of 24 Pearson coefficients was R=0.4808±0.1931, p<0.001 for all 24 linear correlation tests). Additionally, APD was more heterogeneous in highly bumpy surfaces. Fig. 1D and 1E shows that the degree of APD dispersion (SD and range of APD) increased with the higher amplitude and the higher number of curves. Both SD and range of APD had positive correlation with the surface bumpiness (Spearman's rho=0.9809 and 0.9623 respectively, both p-values<0.001).

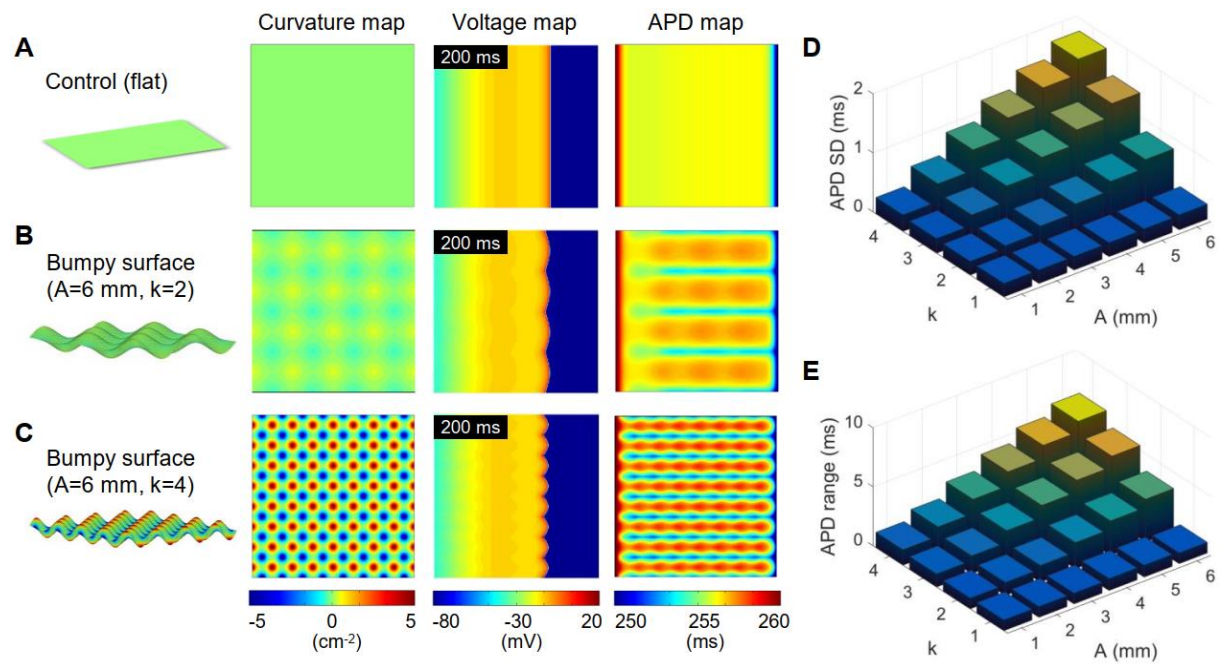

**Fig. 1. Spatial dispersion of action potential duration (APD) in 2D-bumpy atrial tissue at 2 Hz**

**line pacing from left side.** APD was spatially heterogeneous in bumpy surface and linearly correlated to curvature. (A, B, C) Curvature map, voltage map, and APD map are presented for flat control case, bumpy surface of A=6 mm, k=2, and bumpy surface of A=6 mm, k=4, respectively. A and k are the amplitude and the number of curves, respectively. (D, E) The degree of APD dispersion increased as the surface bumpiness increased.

*Formation of wavebreak in bumpy geometry*

We induced single spiral wave and examined whether wavebreak occurred spontaneously. We considered the flat tissue as a control case to compare with bumpy tissue. In the flat control case (Fig. 2A), only single spiral wave existed without any formation of wavebreak until it spontaneously terminated. The PS map with the control case shows only single trajectory of the spiral wave core. However, wavebreaks were observed in the nine surfaces of high amplitude and number of curves (Fig. 2C). As shown in Fig. 2B, wavefront was broken up into multiple wavelets when a part of the wavefront failed to propagate in the region not recovered from refractoriness. The broken ends of wavelets begin to rotate and these reentry centers are shown in the PS map. In the nine surfaces with wavebreak, AF maintenance duration was prolonged compared to that of the flat control case (4.6±0.6 vs. 3.7 s, p=0.004, Fig. S1-B), and surface bumpiness was higher than that in the other 15 surfaces (0.60±0.53 vs. 0.09±0.15 $cm^{-2}$, p=0.001, Fig. 2D). We re-performed all the simulations after standardizing the area to eliminate the effects of the tissue size on fibrillation. The results were consistent to that of the original simulation (Fig. S1-C and D).

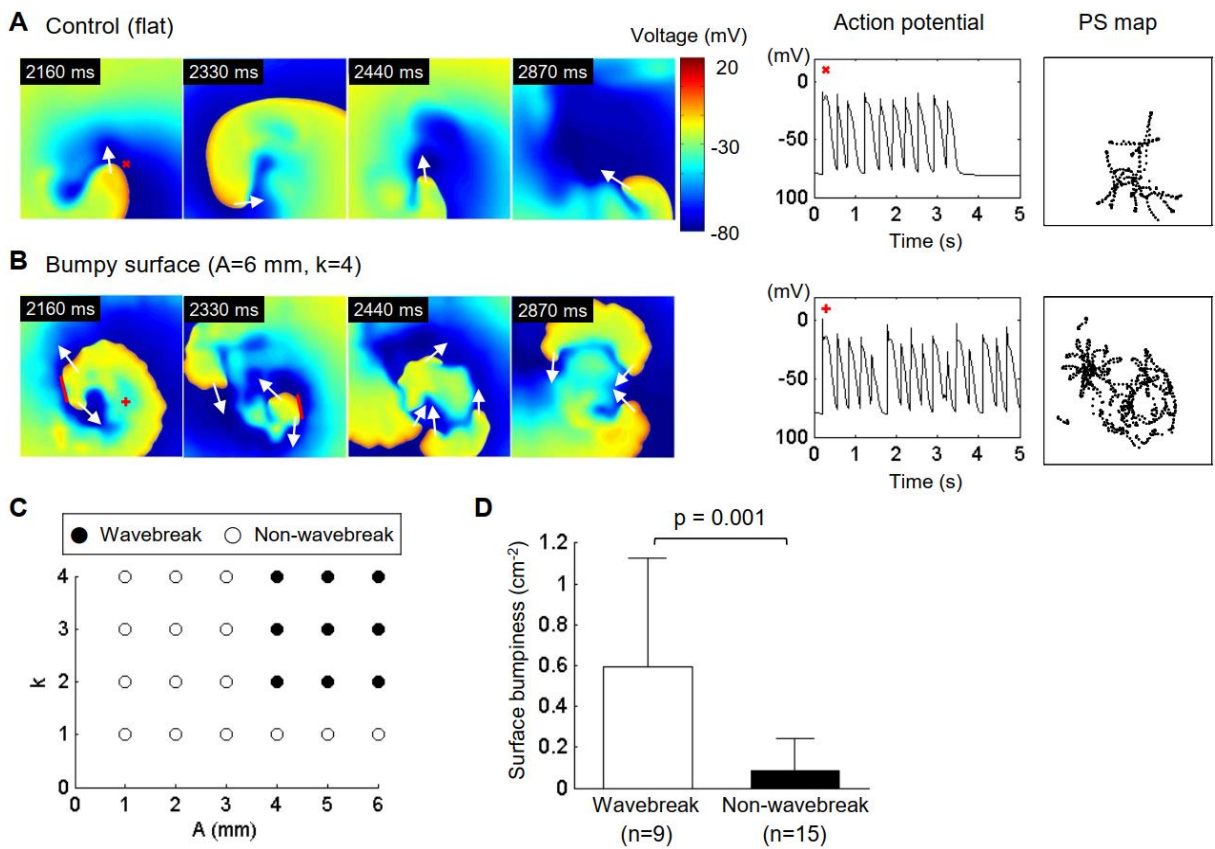

**Fig. 2. Formation of spontaneous wavebreak in bumpy surface.** In 2D-bumpy model, bumpy geometry could induce wavebreak without presence of electrical remodeling. (A, B) Snapshot of wave propagation, action potential, and phase singularity (PS) map are presented for flat control case and bumpy surface of A=6 mm, k=4, respectively. A and k are the amplitude and the number of curves, respectively. White arrows indicate direction of wave propagation and red lines indicate conduction block due to refractoriness. Action potential recording points are marked by red symbols. (C) Spontaneous wavebreak occurred in highly bumpy surfaces (nine black dots). Surfaces with wavebreak are marked by black dots. White dots indicate that only single spiral wave existed without any wavebreak. (D) Comparison of surface bumpiness between surfaces with wavebreak and the other cases (Mann–Whitney U test, p=0.001).

Furthermore, we generated 200 random non-uniformly bumpy surfaces and determined whether wavebreak spontaneously generated and whether AF maintained for 5 seconds in fibrillation state (see Fig. 3). The surfaces with wavebreak had higher surface bumpiness than that in the non-wavebreak cases (p<0.001), and the surface bumpiness of the AF-sustained cases was higher than that of the other cases (p<0.001). The logistic regression model could predict wavebreak generation (OR=934.68 [95% CI 105.18–8306.01], AUC=0.9525, p<0.001) and AF sustainability (OR=45.45 [95% CI 12.28–168.19], AUC=0.9233, p<0.001) from the surface bumpiness value.

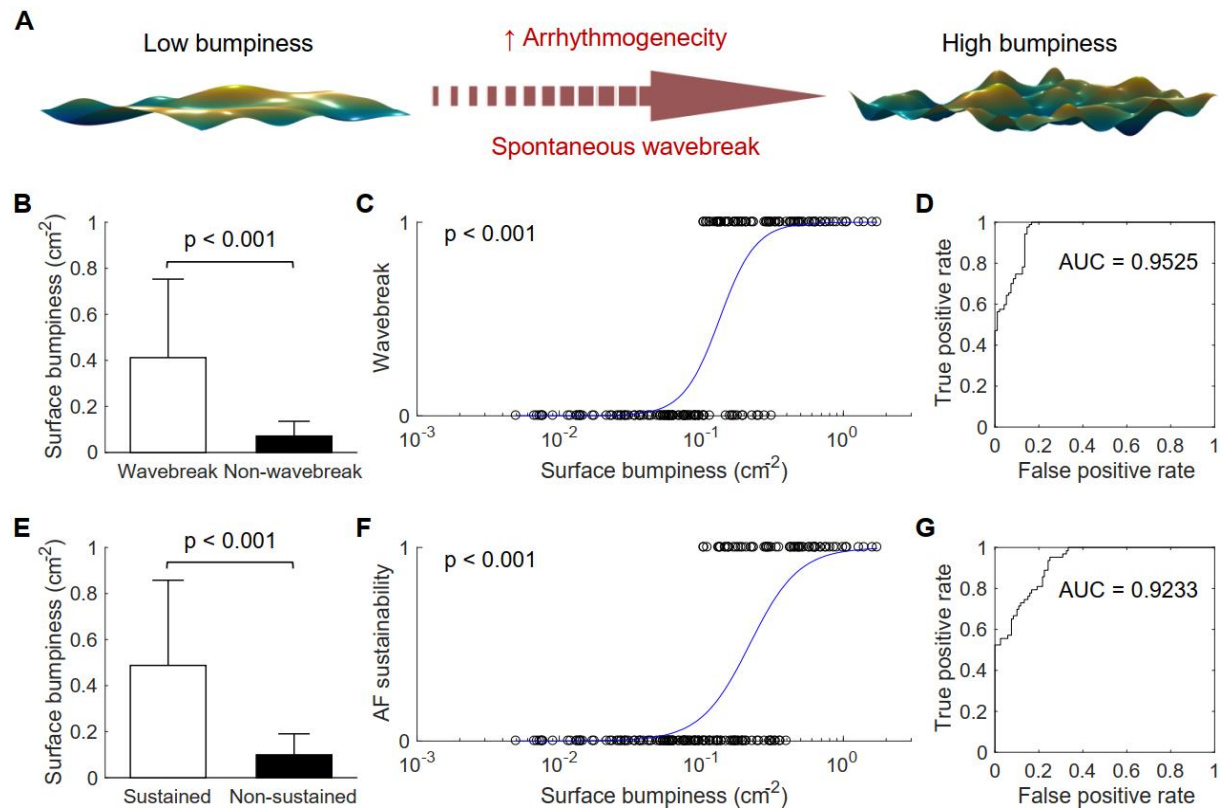

**Fig. 3. Phase transition-like behavior of spiral wave dynamics.** (A) Random non-uniformly bumpy surfaces (n=200) were generated for testing wavebreak generation and AF sustainability (duration>5 s).

(B) Comparison of bumpiness between wavebreak cases and non-wavebreak cases (t-test, p<0.001). (C) Logistic regression analysis between wavebreak generation and bumpiness (p<0.001). (D) Receiver operating characteristic analysis of the logistic regression model (AUC=0.9525). (E) Comparison of bumpiness between AF-sustained cases and non-sustained cases (t-test, p<0.001). (F) Logistic regression analysis between AF sustainability and bumpiness (p<0.001). (G) Receiver operating characteristic analysis of the logistic regression model (AUC=0.9233).

### *Drift of spiral wave against bumpiness gradient*

To determine the effects of heterogeneous bumpiness on spiral wave dynamics, we simulated non-uniformly bumpy surface with bumpiness gradient. In the uniformly bumpy case (Fig. 4A), the spiral wave sustained for more than 5 s with wavebreaks, and more PS points were located at left half plane than right half plane (55.1% vs. 44.9%). However, in the surfaces with bumpiness gradient (Fig. 4B-D), more than 85% of PS points were located at right half plane, which had higher bumpiness compared with left half plane (85.7%, 96.8%, 97.4% for α=1, 2, 3, respectively). The spiral wave drifted towards the area of higher surface bumpiness.

The combined gradient of bumpiness and ionic current showed that PS points were concentrated in the higher bumpy area and wavebreak occurred spontaneously; however, the spiral wave drift was not prominent, which might have been masked by the electrical gradient (Fig. S2).

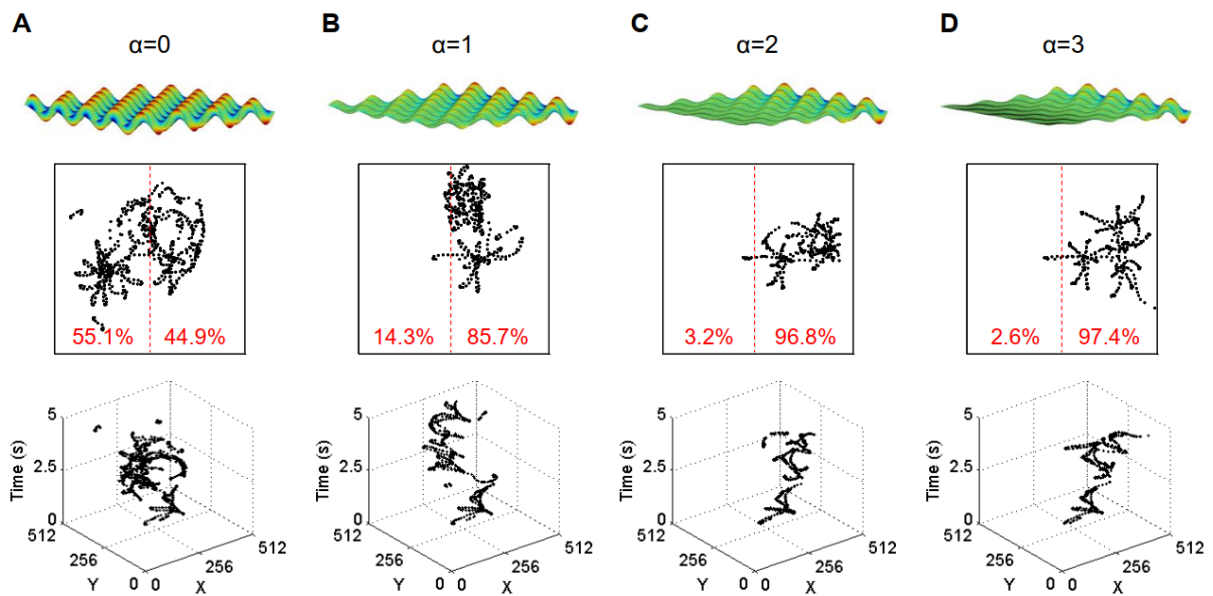

**Fig. 4. Spiral wave drift and wavebreak in bumpy surface with bumpiness gradient.** In 2D-bumpy model with bumpiness gradient, spiral wave drifted towards highly bumpy region (right side). Phase singularity (PS) maps (second row) and PS trajectories in space-time domain (third row) are presented for four different degrees of bumpiness gradient (α=0, 1, 2, 3). Portion of PS points in the right half side increased as the degree of bumpiness gradient increased (red numbers, %).

## 3D human AF simulation

To determine the formation of spontaneous wavebreak in the LAA, we simulated 3D human AF model reflecting the LA anatomy of 14 AF patients. In all the cases, AF was maintained for longer than 6 seconds and wavebreak was observed in the entire LA including the LAA. Since PS is known to be closely related to wavebreak,[24] we generated the PS map for 6 seconds in AF state. In both isotropic and anisotropic models, the PS density in the LAA was significantly higher than that in the other LA parts (see Fig. 5).

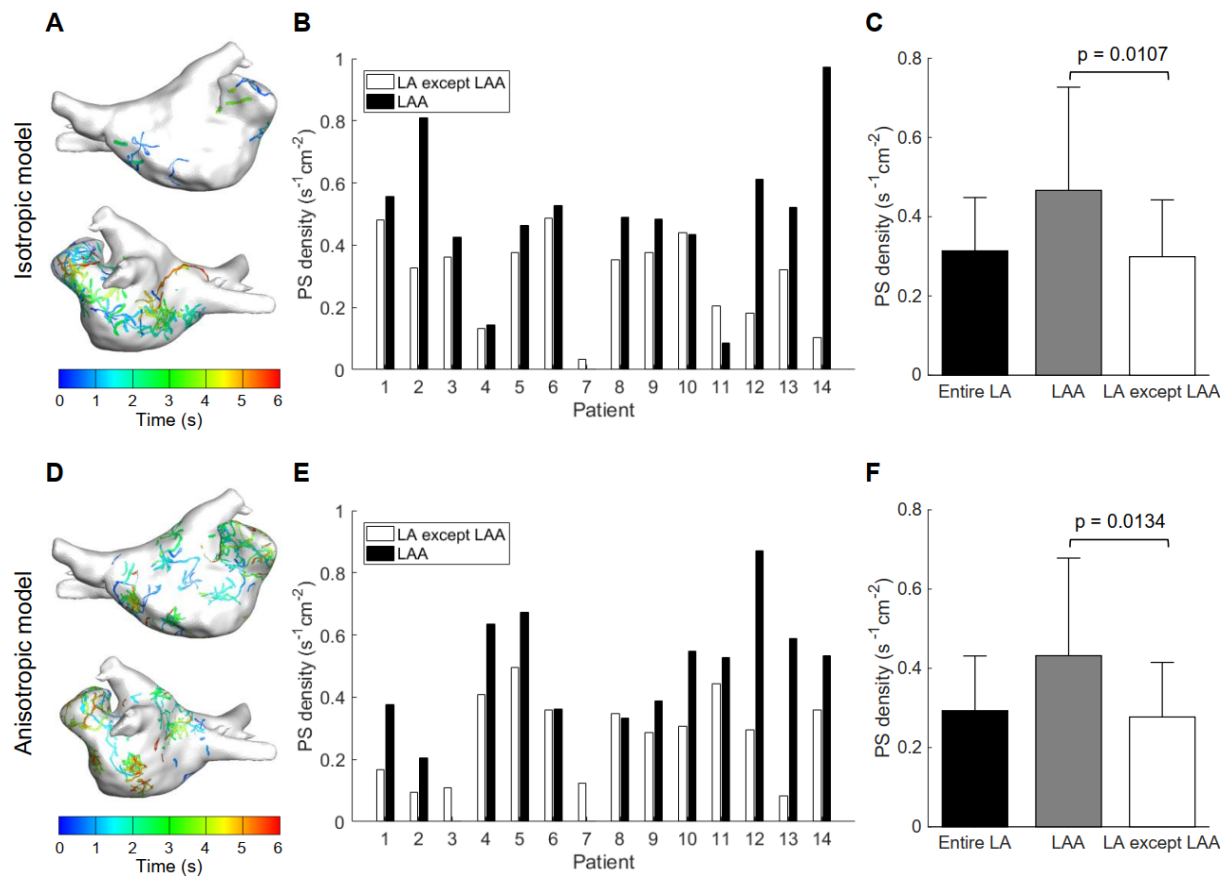

**Fig. 5. Phase singularity (PS) map in 3D left atrial (LA) model (n=14).** In *in silico* 3D LA model of atrial fibrillation (AF), AF was successfully induced and wavebreak occurred. (A) Example of PS map in isotopic model. (B, C) Comparison of PS density between LAA and other LA parts in isotropic model (Wilcoxon signed-rank test, p=0.0107). (D) Example of PS map in anisotropic model. (E, F) Comparison of PS density between LAA and other LA parts in anisotropic model reflecting fiber orientation (Wilcoxon signed-rank test, p=0.0134).

## Clinical evidences for arrhythmogenic LAA
### *Curvature map of LA geometry*

In 30 patients with persistent AF, we measured the curvature of LA geometry reconstructed from CT images. Fig. 6A shows that the curvature of LA geometry was heterogeneous over the entire LA. The surface bumpiness in the LAA was 5.8-fold compared with the other LA parts (1.52±0.33 vs. 0.26±0.04

cm$^{-2}$, p<0.001, Fig. 6E). All bumpiness values of the LAA had more than 99% probability of wavebreak generation according to the logistic regression model developed from the 2D simulation (see Fig. 3).

*Wave dynamics parameters of bipolar electrogram*

We clinically obtained CFAE-CL, ShEn, and ApEn maps of the same group of 30 patients with contact bipolar electrograms taken during clinical procedures (Fig. 6B-D). Both parameters were utilized for evaluating the complexity of fractionated electrogram, which is known to be closely related to wavebreak [14]. The CFAE-CL in the LAA was lower than that of the other LA parts (130.9±29.4 vs. 154.7±32.2 ms, p<0.001), and the percentage area of CFAE was higher in the LAA than that in the other LA parts (40.1±22.0 vs. 30.1±14.5%, p<0.001, Fig. 6F). Moreover, there was an inverse relationship between the surface bumpiness of the LAA and the mean CFAE-CL of the entire LA (R=-0.376, p=0.041). Additionally, both ShEn and ApEn in the LAA were higher compared to those in the other LA parts (ShEn: 4.40±0.66 vs. 3.50±0.39; ApEn: 0.10±0.04 vs. 0.04±0.02; both p<0.001, Fig. 6G).

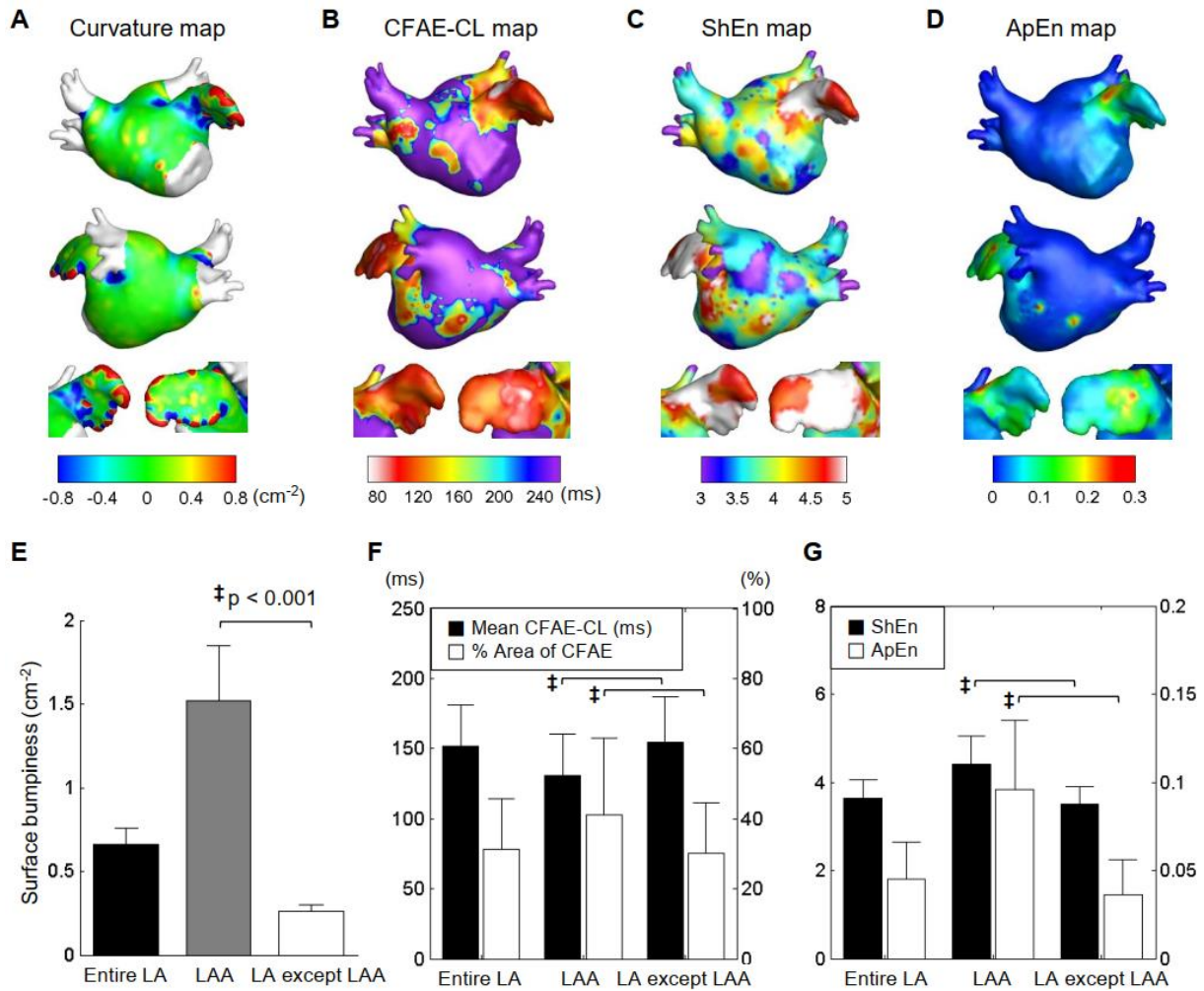

**Fig. 6. The 3D left atrial (LA) mapping of curvature, complex fractionated atrial electrogram-cycle length (CFAE-CL), Shannon entropy (ShEn), and approximate entropy (ApEn) for patients with persistent AF (n=30).** Wave dynamics parameters were calculated from clinically-recorded

bipolar electrogram. All parameters were dominant in LA appendage (LAA). (A, B, C, D) Example of curvature, CFAE-CL, ShEn, and ApEn map, respectively. Anterior and posterior parts of LA and LAA were displayed. (E) Surface bumpiness in LAA and other LA parts (paired t-test, p<0.001). (F) Mean CFAE-CL and percentage area of CFAE (CFAE-CL≤120ms) in LAA and other LA parts. (G) Entropy parameters (ApEn and ShEn) in LAA and other LA parts (paired t-test, ‡p<0.001).

## Discussion

We discovered the arrhythmogenic role of bumpy tissue geometry using computational model of human atrial tissue. In 2D-bumpy surface model, the bumpy geometry induced APD dispersion, which led to formation of wavebreak at high surface bumpiness. Additionally, heterogeneous bumpiness caused spiral wave drift towards the region with higher bumpiness.

In 3D-LA geometry, LAA showed significantly high surface bumpiness, which produced spontaneous wavebreak in 2D-bumpy model. We performed 3D AF simulation and analyzed clinically obtained wave dynamics parameters. Both simulation and clinical data showed that the wave dynamics parameters were significantly higher in LAA compared with other LA parts.

*Effects of bumpy geometry on spiral wave dynamics*

The role of curvature of tissue geometry in electrical wave dynamics was primary discovered by Davydov and Zykov[11]. They theoretically explained the drift of spiral waves on non-uniformly curved surface. Several computational and experimental studies showed that tissue shape, structural heterogeneity, boundary features, and activation time could induce heterogeneity of refractoriness,[9, 10] which has been known to play an important role in initiation of reentry[25] and generation of wavebreaks.[13, 26] We discovered the quantitative relationship between APD heterogeneity and curvature of tissue geometry, and its role in wavebreak generation in bumpy surface. Additionally, we found that the bumpy tissue geometry can localize PS points by attracting the spiral wave against the bumpiness gradient. This novel finding supports that not only ionic current heterogeneity,[27, 28] but also structural heterogeneity contributes to spiral wave drift.[29] Therefore, high surface bumpiness of tissue geometry may be a potential source of arrhythmia.

The effect of curvature of tissue geometry can be mathematically explained by the diffusion equation of electrical conduction[30]. Under the assumption of isotropic conduction, the diffusion term can be represented by inverse metric of the surface as $D \sum g^{ij} \partial_i \partial_j V$ where D is the diffusion coefficient and $g^{ij}$ is the inverse metric of surface. If we define "adjusted diffusion coefficient" as $\widetilde{D}_{ij} = D \cdot g^{ij}$, the diffusion term can be expressed as $\sum \widetilde{D}_{ij} \partial_i \partial_j V$. Therefore, heterogeneous curvature is mathematically equivalent to heterogeneous and anisotropic conduction, which are already known to be important arrhythmogenic factors in cardiac fibrillation.[26]

*Roles of the LAA in AF maintenance*

According to the study of Di Biase et al[2], 27% patients undergoing redo ablation showed the LAA as the trigger site of AF. Recently, they reported results of 24-month follow-up the BELIEF trial showing that the LAA isolation improved the success rate of AF treatment in the patients with long-standing persistent AF.[4] There are several potential mechanisms describing the arrhythmogenic role of the LAA.[7] Since embryological origin of the LAA is mesodermal myocardium which is related to the opening of the pulmonary vein, the LAA may similarly act as the PVs in the initiation of AF. Another anatomic consideration is the ligament of Marshall located between the LAA and the left superior PV. It contains sympathetic and parasympathetic nerves, which can form a triggering source of AF. Additionally, complex fiber orientations of the LAA may influence wave propagation and formation of localized reentry in the LAA.[3]

We proposed another potential mechanism for demonstrating that the bumpy morphology of the LAA may contribute arrhythmogenesis of AF. As shown in the Fig. 3, we observed critical phase transition-like phenomenon, which implies that extremely bumpy structure such as the LAA has potential to generate wavebreak and to maintain AF. The simulation result and the clinical data indicated the present of wavebreak or spiral wave reentry in the LAA. According to previous studies, CFAE, ShEn, and ApEn indicate the complexity of fractionated electrogram, which is closely related to wavebreak and arrhythmogenic characteristics of AF.[14-16] Additionally, the high entropy parameters are known to be associated with rotational activation called 'rotor',[15, 16] which is usually produced by wavebreak at the broken ends of wavelets[24]. Therefore, overall results of our study imply that the bumpy morphology of the LAA may play a pro-arrhythmic role in human AF.

*Clinical implications*

Although many clinical evidences have emerged to support the efficacy of the LAA isolation in controlling AF[2-6], the mechanism of the LAA in the maintenance of AF is still unrevealed. We proposed a potential mechanism that the bumpy morphology of the LAA may play a pro-arrhythmic role in AF. Of course, the tissue heterogeneity and autonomic neural activity significantly contribute to the initiation and maintenance mechanisms of arrhythmia. However, we found that anatomical bumpiness of the cardiac structures also contributes to the APD dispersion and arrhythmogenesis. Additionally, local bumpiness is linearly correlated to LA wall thickness, which is closely associated with local wave dynamics parameters.[31] Therefore, not only the evaluation of the tissue heterogeneity but also the measurement of the surface bumpiness may contribute to the accurate mapping of potential target of arrhythmia in ablation procedure.

*Limitations*

In 2D simulation, we only considered periodic bumpy pattern without testing global structural heterogeneity of LAA, which must play a role in atrial arrhythmogenicity. Also, we simulated 12.8×12.8

cm$^2$ tissue to avoid the issue of critical mass hypothesis.[32, 33] However, quantitative results in 2D simulation cannot be directly compared to either simulation results of 3D-LA model or real electrical wave dynamics in patients with AF, due to the topological discrepancy such as existence of tissue boundary. In our 3D AF model, we did not consider regional LA wall thickness, tissue heterogeneity, and autonomic neural activity to test the exclusive roles of bumpy tissue morphology and to reduce the computational cost, despite of the increasing role of intramural conduction in AF.[34] The sample size of 3D-patient specific modeling was relatively small, which could have affected the results. However, we confirmed reproducible results in both parametric and non-parametric analyses. Due to technical limitations in accessing clinical APD and PS data from patients with AF, we considered CFAE-CL and ShEn as parameters for estimating electrical wave dynamics, however, CFAE-CL also reflects sequentially acquired electrograms with low spatial resolution, far-field potential, overlapped myocardial fibers, and AF drivers[14]. Complex nonlinear relationship between structural heterogeneity and wave dynamics should be further investigated.

## Conclusion

In the computational model of human atrial tissue, bumpy tissue morphology induced APD dispersion, which promoted formation of spontaneous wavebreak. Additionally, bumpiness gradient caused spiral wave drift in the direction of higher bumpy area. In 3D-LA geometry, surface bumpiness of the LAA is significantly higher than the other LA parts. Both simulation results and clinical data showed that wave dynamics parameters were significantly higher in the LAA than those in the other LA parts, implying potential arrhythmogenic mechanism of the LAA in human AF.

## Acknowledgements

This work was supported by a grant [A085136] from the Korea Health 21 R&D Project, Ministry of Health and Welfare and a grant [NRF-2017R1A2B4003983] from the Basic Science Research Program run by the National Research Foundation of Korea (NRF) which is funded by the Ministry of Science, ICT & Future Planning (MSIP).